\documentclass[a4paper,preprint,superscriptaddress]{revtex4-1}
\usepackage{amsmath}
\usepackage{amssymb}
\usepackage{graphicx}
\usepackage{bm}
\usepackage{url}
\begin{document}
\title{Enhanced Sampling in the Well-Tempered Ensemble}
\author{Massimiliano Bonomi}
\email{mbonomi@ethz.ch}
\affiliation{Computational Science, Department of Chemistry and Applied Biosciences, ETH Zurich, USI Campus, via G. Buffi 13, 6900 Lugano, Switzerland}
\author{Michele Parrinello}
\email{parrinello@phys.chem.ethz.ch}
\affiliation{Computational Science, Department of Chemistry and Applied Biosciences, ETH Zurich, USI Campus, via G. Buffi 13, 6900 Lugano, Switzerland}
\date{\today}
\begin{abstract}
We introduce the well-tempered ensemble (WTE) which is the biased ensemble
sampled by well-tempered metadynamics when the energy is used as collective variable.
 WTE can be designed so as to have approximately the same average energy as the
canonical ensemble but much larger fluctuations.  These two properties lead to an
extremely fast exploration of phase space. An even greater  efficiency is obtained
when WTE is combined  with parallel tempering.  Unbiased Boltzmann averages are
computed on the fly by a recently developed reweighting method
\small{[M. Bonomi \emph{et al.}  J. Comput. Chem. {\bf 30}, 1615 (2009)]}.
We apply WTE and its parallel tempering variant to the 2d Ising model and to a G\={o}-model of HIV
protease,  demonstrating in these two representative cases that convergence is accelerated by orders of magnitude.
\end{abstract}
\maketitle
Monte Carlo (MC) or molecular dynamics (MD) simulations are 
routinely applied in all areas of science.
However, severe difficulties  are encountered when
multiple metastable states separated by large free-energy barriers are present.  
Nucleation from one phase to another, chemical reactions, and protein folding are important examples. 
Accessing the low probability regions separating one state from another can overcome this difficulty. 
In standard MC or MD this is not possible and the system remains confined to its initial basin 
hindering a proper phase space exploration.
Sampling low probability regions would also be of great help in free-energy 
differences calculation \cite{frenkelsmit}.
Hence many enhanced sampling methods have been suggested 
\cite{hans97cpl,sugi-okam99cpl,mari-pari92el,fuku+02jcp,wanglandau,
umbrella_sampling,flooding,willy,jarzynski,pohorille,tps2}.

Recently, we have developed metadynamics \cite{metad} where 
few difficult to sample degrees of freedom or collective variables (CV) are
selected \cite{eppur,Laio:2008p11564}. 
If the CV are well chosen large free-energy barriers 
can be overcome and the associated free-energy surface (FES) reconstructed \cite{bussi_noneq}.  
Well-tempered metadynamics \cite{Barducci:2008} is a non-trivial evolution of the method
which lends itself to reweighting thus 
allowing the calculation of unbiased canonical averages \cite{Bonomi:2009p17105}.  
We show here that when the potential energy is used as CV 
a well definite distribution dubbed well-tempered ensemble (WTE) is sampled.
Using WTE is possible to observe transitions between 
states that otherwise would have been impossible to study in standard MC or MD. 

Many approaches have been already suggested in which the energy distribution is altered
artificially \cite{PhysRevLett.92.170601,Michel:2009p17713,li:094101,zheng:014105,Donadio:2005p17962,CHALLA:1988p18009}.
However, all these methods can evaluate only the density of states from which thermal properties
can be determined. If information on other variables is needed for each new variable
a separate calculation is required \cite{Michel:2009p17713,NEUHAUS:2006p18010}. 
Here instead full information on all the variables distribution can be obtained from a single run.
Furthermore in an appropriate combination with 
parallel tempering (PT) \cite{bussi_xc},
we show that orders of magnitude sampling efficiency can be gained.

Let us use as CV the potential energy $U=U(\bm{R})$
where $\bm{R}$ is the full set of atomic coordinates.
In well-tempered metadynamics the Newton's equations are altered by the addition
of a bias potential $V(U(\bm{R}),t)$:
\begin{equation}
\bm{m}\ddot{\bm{R}}=-\frac{\partial U(\bm{R})}{\partial \bm{R}}-\frac{\partial V(U(\bm{R}),t)}{\partial \bm{R}},
\label{newton}
\end{equation}
whose time evolution is governed by:
\begin{equation}
\dot{V}(U,t)=\omega e^{-\frac{V(U,t)}{k_B \Delta T}}\delta_{U,U(t)},
\end{equation}
where $\bm{m}$ are the atomic masses, while $\omega$ and $\Delta T$ are 
parameters which have the dimension of an energy rate and a temperature respectively. 
Asymptotically,  $V(U,t)$:
\begin{equation}
V(U,t\rightarrow \infty)=-\left (1-\gamma ^{-1} \right) F(U),
\end{equation}
with $\gamma=(T+\Delta T)/T \geq 1$ and $F(U)=-\frac{1}{\beta} \ln{\frac{\int{d\bm{R} \, \delta (U-U(\bm{R})) \,e^{-\beta U(\bm{R})}}}{\int{d\bm{R} \,e^{-\beta U(\bm{R})}}}}$. 
Within an irrelevant constant,
\begin{equation}
F(U) = U-\beta^{-1}\ln{N(U)}
\end{equation}
where $N(U)=\int{d\bm{R} \, \delta (U-U(\bm{R}))}$ is the number of states 
of energy $U$, which is a $T$ independent property \cite{PhysRevLett.92.170601,Michel:2009p17713,wanglandau}. 
$V(U,t)$ quickly converges to its $t \rightarrow \infty$ limit and 
the configurations are distributed according to:
\begin{equation}
\mathcal{Z}_\gamma=\int{d\bm{R}\,e^{-\beta U_\gamma(\bm{R})}},
\label{partition_WTE}
\end{equation}
with
\begin{equation}
U_\gamma (\bm{R})=U(\bm{R})-\left( 1-\gamma^{-1} \right) \left [  U(\bm{R})  - \beta^{-1} \ln{N(U(\bm{R}))} \right ],
\end{equation}
which defines WTE. It is then easy to rewrite the partition function $\mathcal{Z}_\gamma$ as:
\begin{equation}
\mathcal{Z}_\gamma=\int{dU\, P(U)^{\frac{1}{\gamma}}},
\end{equation}
where $P(U)=e^{-\beta U} N(U)$  is proportional to the energy probability distribution in the canonical ensemble. 
Varying $\gamma$ one goes from the canonical partition function ($\gamma=1$) to the multicanonical one ($\gamma=\infty$)
 \cite{BERG:1991p17843}.
In order to gain insight into the $\mathcal{Z}_\gamma$ properties we make the assumption that
$P(U)$ is strictly Gaussian,  $P(U) \propto e^{-\frac{(U-\langle U \rangle)^2}{2\Delta U^2}}$,
where $\langle U \rangle$ is the average energy in the canonical ensemble and $\Delta U^2$ the corresponding 
fluctuation \cite{AMADEI:1996p17722}.
Thus $P(U)^{\frac{1}{\gamma}} \propto e^{-\frac{(U-\langle U \rangle)^2}{2\gamma \Delta U^2}}$
 implying the same average energy as in the canonical case $\langle U \rangle_\gamma=\langle U \rangle$ but $\gamma$ time larger fluctuations.
The Gaussian assumption is not always 
justified, for instance when $\langle U \rangle$ is 
close to the ground state energy or to a critical point. 
Still for reasonably large $\gamma$ one finds 
$\langle U \rangle_\gamma$  close to $\langle U \rangle$ with fluctuations which grow approximately linearly with  $\gamma$. 
In a rather loose sense it is as if  a quasi-critical behavior is induced at all temperatures. 
 This similitude is further increased by the fact that dynamical correlations are slowed down.
However, when $\gamma$ increases even further the non-Gaussian 
tails in $P(U)$  are amplified until for $\gamma \rightarrow \infty$ one reaches the multicanonical limit.

We now combine WTE with PT (PT-WTE). 
In PT,  $n$  replicas of the system at the temperatures $\beta_i, \, i=1,n$ are introduced and a MC procedure is used
to attempt exchanging  configurations between replicas.  Colder replicas are prevented from being trapped in 
local minima by the exchange with the  higher temperature ones. 
A figure of merit is the ability of a  replica to 
diffuse across all range of $\beta_i$ and 
methods that speed up this diffusion have been suggested (see Ref.~\cite{Earl:2005p16718} and references within).
Given the special properties of WTE, it is tempting to explore its performance when combined with PT
since one expects that the enhanced energy fluctuations will greatly facilitate exchange processes.
In addition, if one use the same $\gamma$ factor for all the $\beta_i$,
 the swapping probability in PT-WTE is determined by:
\begin{equation}
\Delta_{i,j}=\gamma^{-1}\,(\beta_i-\beta_j)(U(\bm{R}_i)-U(\bm{R}_j)),
\label{WTE_delta}
\end{equation}
 implying a factor $\gamma$ reduction relative to conventional PT ($\gamma=1$).
This is possibly the main result of this paper and shows why PT-WTE leads to fast diffusion  across the $\beta_i$.

We now present two representative  applications of WTE and of PT-WTE  to substantiate our claim. 
First we consider the performance of WTE in the single replica mode. 
We simulate the two dimensional ferromagnetic Ising model for which an exact solution exists \cite{FERDINANAE:1969p17790} and 
on which a large number of methods have been tested \cite{Katzgraber:2006p17781,Bittner:2008p17792}. 
The Hamiltonian for this model is: 
$\mathcal{H}=-J\sum_{<i,j>}{S_iS_j}$.
We put $J=1$ and $S_i=\pm1$ are spins on a square lattice with side $L$.
Periodic boundary conditions are applied and only first-neighbor interactions are included.
\begin{figure}[!h]
\begin{center}
\includegraphics[clip,width=1.0\linewidth]{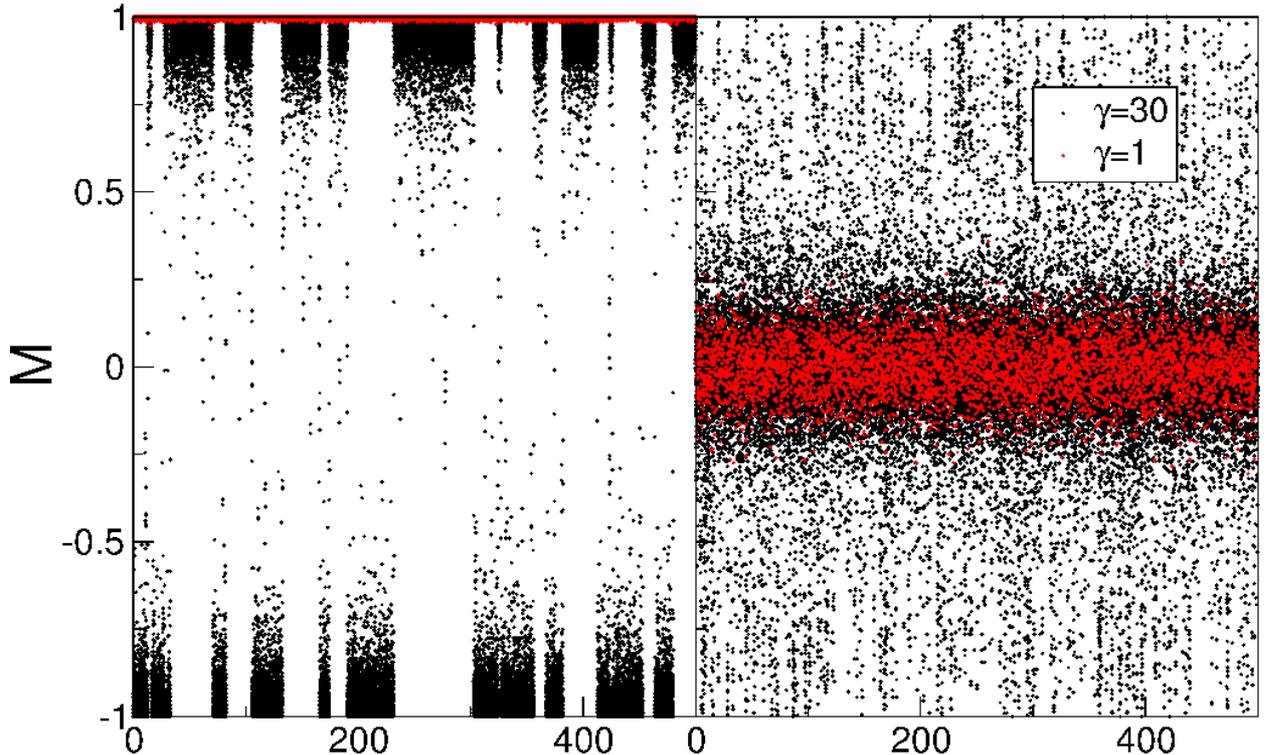}
\end{center}
\caption{WTE (black) compared to standard ensemble sampling (red) at two 
temperatures, below ($T_1=1.0$) and above ($T_2=5.0$) the critical temperature $T_c= 2.269$. 
The unit on the x-axis is $10^3$ MC steps. Each MC move consists of a complete sweep of the L=20 site lattice.  
Gaussians of 0.1 height and 5.0 width were deposited at each step.}
\label{wte_vs_boltz}
\end{figure}
In the ferromagnetic state standard MC explores only one magnetization direction (Fig~\ref{wte_vs_boltz}). 
WTE instead  is able to sample either spin orientations overcoming the large free-energy barrier
($\simeq 110 k_BT$)  that separates these two equivalent states. 
It is also seen that while the average values of the magnetization is approximately correct ($|M|\approx 1 $ in the ferromagnetic 
phase and $M \approx 0$ in the paramagnetic one), the energy fluctuations grow with $\gamma$ (see Table \ref{WTE_table}).
\begin{table}
\begin{center}
\begin{tabular}{c|c|c|c|c|c|c|}
  \cline{2-7} 
    & \multicolumn{3}{c|}{$T_1 < T_c$} & \multicolumn{3}{c|}{$T_2 > T_c$} \\
  \hline
  \multicolumn{1}{|c|}{$\gamma$} & $\langle U \rangle_\gamma $ & $\Delta U_\gamma^2 / \gamma$ & $\tau / \gamma$ &  $ \langle U \rangle_\gamma$ & $\Delta U_\gamma^2 / \gamma$  & $\tau / \gamma$\\
  \hline
  \multicolumn{1}{|c|}{1}    & -798.9 & 9.3    & - & -170.8 & 974.7   &  1.5  \\
  \multicolumn{1}{|c|}{5}    & -790.2 & 31.5   &  0.26 & -174.8 & 999.4   &  1.67 \\
  \multicolumn{1}{|c|}{10}   & -780.1 & 49.8   &  0.23 & -180.2 & 1027.6  &  2.13 \\
  \multicolumn{1}{|c|}{50}   & -710.4 & 154.2  &  0.45 & -206.8 & 1079.5  &  2.58 \\
  \multicolumn{1}{|c|}{100}  & -637.8 & 223.4  &  0.52 & -192.5 & 923.2   &  2.09 \\
  \multicolumn{1}{|c|}{1000} & -193.6 & 180.2  &  0.26 & -39.8  & 199.9   &  0.26 \\
  \hline
\end{tabular}
\end{center}
\caption{Average value, fluctuation and correlation time of the energy in WTE as a function
of $\gamma$ at the two representative temperatures, below and above $T_c$.
The value of $\tau/\gamma$ at $\gamma=1$, $T=T_1$ is smaller than a single sweep.} 
\label{WTE_table}
\end{table}
For $T\!\!>\!\!T_c$ the Gaussian assumption is clearly justified since $\langle U \rangle_\gamma$ and $\Delta U_\gamma^2 / \gamma$ are approximately 
constant up to $\gamma \sim 100$. 
For $T\!\!<\!\!T_c$ and up to $\gamma \sim 100$, $\langle U \rangle_\gamma $ is also little shifted. However, 
the non linear fluctuation growth signals deviations from Gaussian behavior due to the proximity to the energy  
lower bound. In both cases relaxation times grow linearly with $\gamma$ and do not outweigh the benefit of increased fluctuations. 
We expect a useful $\gamma$ to be of the order of $\gamma \simeq k_B T \Delta F / \Delta U^2$,
where $\Delta F$ is the relevant barrier. As such, $\gamma$ will be system and size dependent.

\begin{figure}[!h]
\begin{center}
\includegraphics[clip,width=1.0\linewidth]{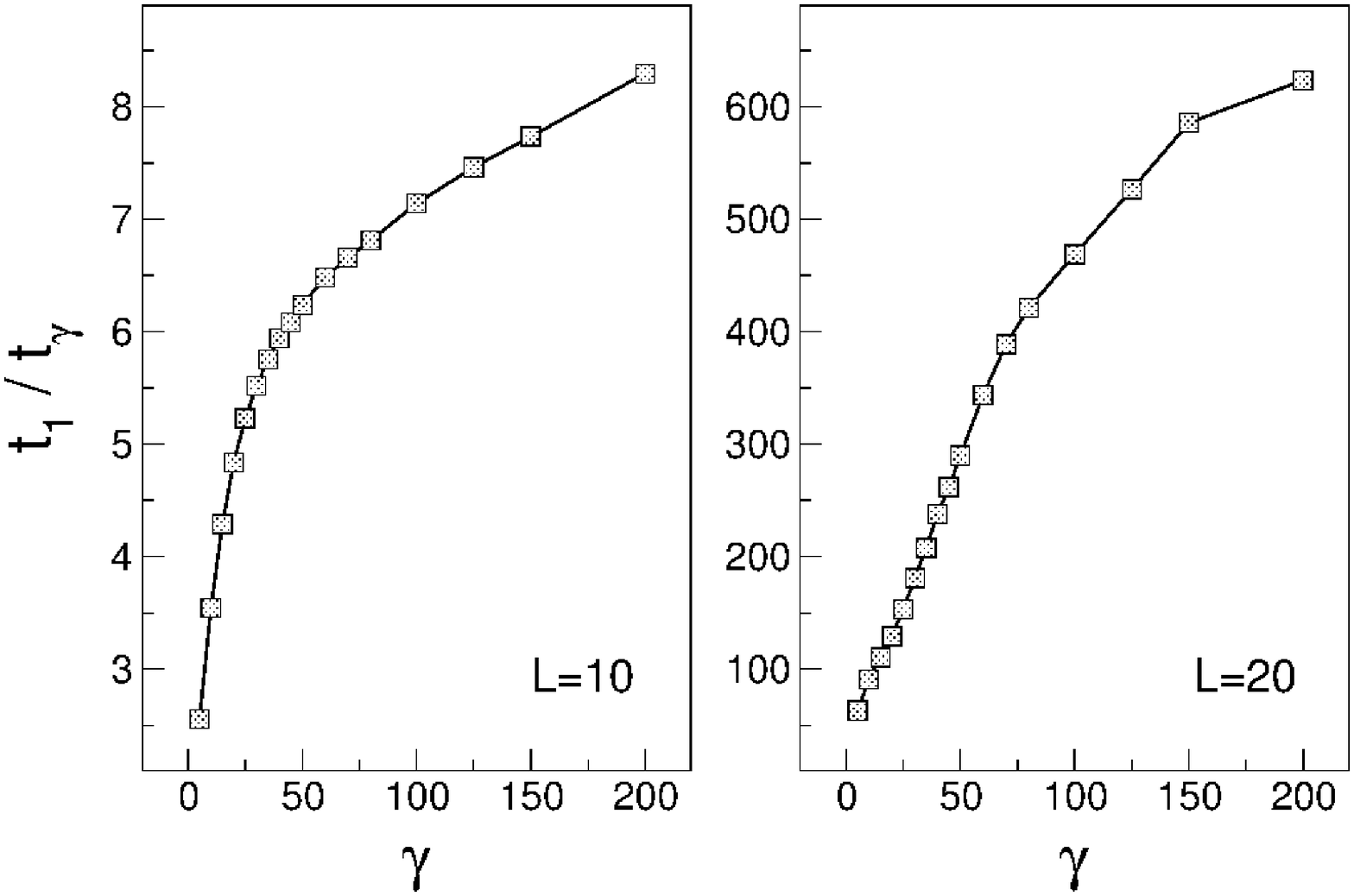}
\end{center} 
\caption{Speed-up of PT-WTE compared to standard PT as a function of $\gamma$ in the
Ising model with L=10 (left panel) and L=20 (right panel).
21 replicas were distributed in a geometric progression in the
interval $0.1\!\!\le\!\!T\!\!\le\!\!10.0$ as in Ref.~\cite{Katzgraber:2006p17781}.
Exchange moves were attempted after every lattice sweep. 
Gaussian parameters as in Fig.~\ref{wte_vs_boltz}.}
\label{speed-up}
\end{figure}

Despite the fact that we have not attempted to optimize the replica distribution \cite{Katzgraber:2006p17781}, 
the use of WTE leads to a great improvement in efficiency when combined with PT.
This is measured in terms of round-trip time $t_\gamma$, which is the
time needed for a configuration in the coldest replica to reach the hottest temperature and come back \cite{Katzgraber:2006p17781}.
It can be seen in Fig.~\ref{speed-up} that the speed-up grows  almost linearly with $\gamma$
up to $\gamma \simeq 30$ for $L=10$ and $\gamma \simeq 100$ for $L=20$, and is much larger than what reported by optimizing the $\beta_i$
distribution \cite{Katzgraber:2006p17781}.
Empirically, the ratio between the smallest energy difference between successive $\beta_i$
and the largest energy fluctuation measured in the unbiased ensemble provides a good estimate for the
optimal $\gamma$.
Above this value the speed-up ceases to be linear in $\gamma$ and the increased fluctuations
and the reduction in acceptance ratio do not compensate the dynamical slowing down.
\begin{figure}[!h]
\begin{center}
\includegraphics[clip,width=1.0\linewidth]{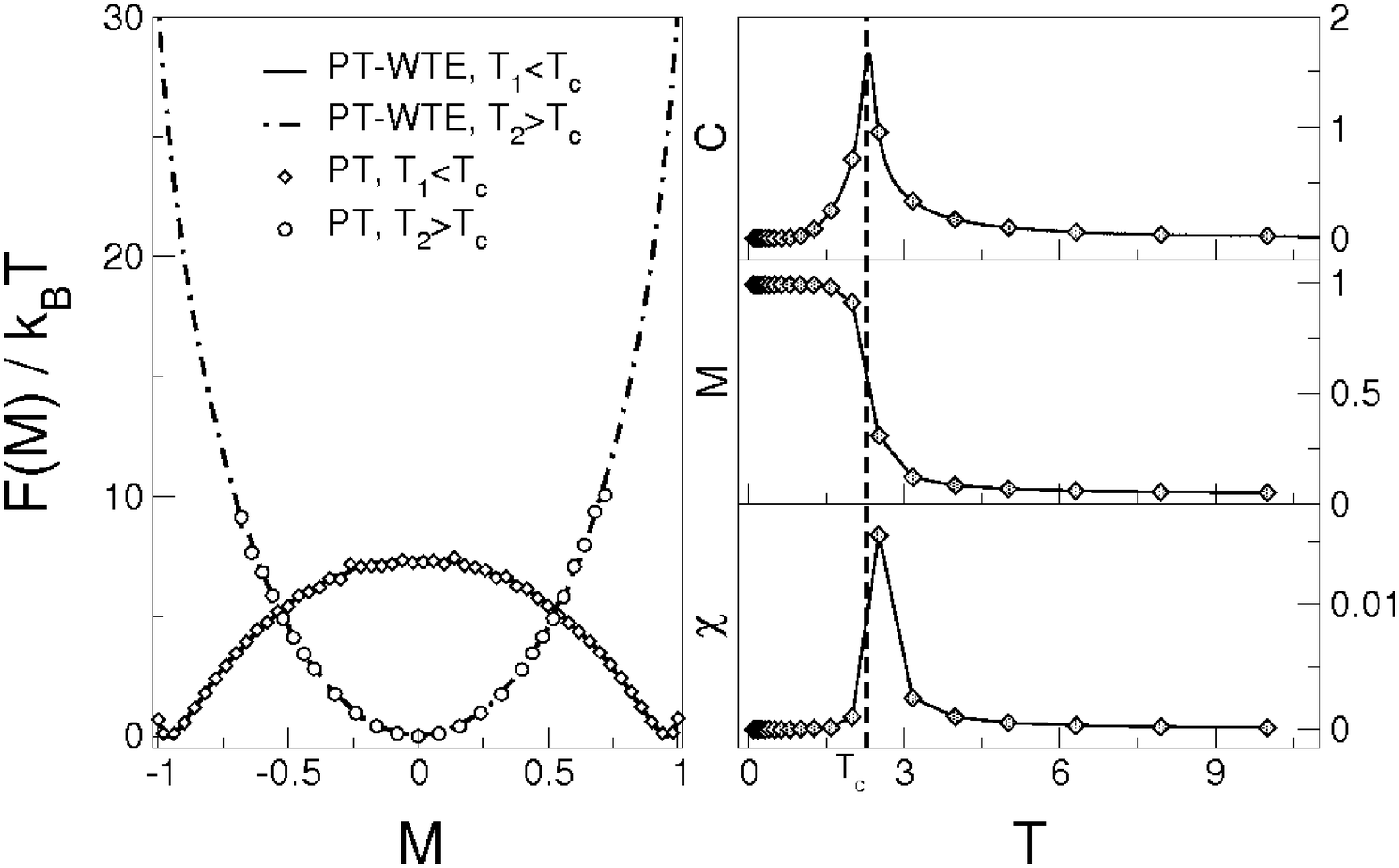}
\end{center}
\caption{Left panel. FES as a function of the magnetization $F(M)$ of the L=10 next neighbor ferromagnetic Ising model below and above
the critical temperature, compared with an extensive PT calculation. 
The statistical error for the PT-WTE calculations is smaller than 2 \%. 
We have computed a similar curve for L=20  but we do not show it here 
because the PT calculation to compare with could not be converged. 
It is remarkable that both magnetization could be explored in spite of a barrier of the order of $110 k_BT$.
 Right panels. Specific heat per spin (top), modulus of the magnetization (middle) 
and magnetic susceptibility (bottom) as a function of temperature (L=20).
The continuos line in the top panel is the finite size exact solution \cite{FERDINANAE:1969p17790}.
In the middle and bottom panel the line is just a guide to the eye.
The statistical error found is at worst ~1 \% in all cases.
} 
\label{F_M}
\end{figure}

\begin{figure}[!h]
\begin{center}
\includegraphics[clip,width=1.0\linewidth]{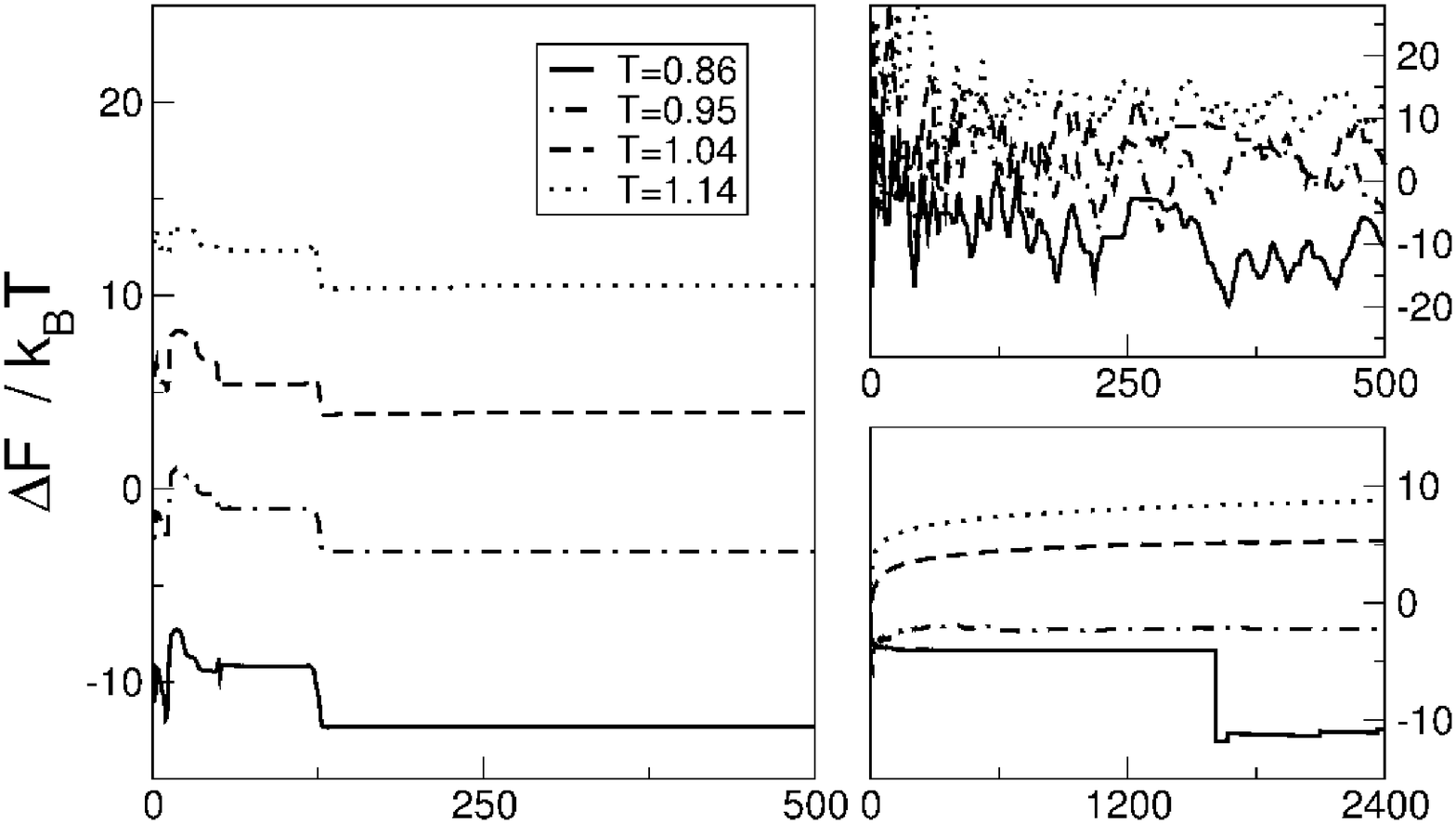}
\end{center}
\caption{Left panel. G\={o}-model FES convergence in PT-WTE run measured as the free-energy difference
between folded and unfolded state as a function of time in units of $10^5$ MD steps. 
16 replicas were distributed with a geometric progression in a temperature range between 0.625 and 1.25
in unit of $T_f$.
Exchanges between configurations were attempted every 200 MD steps.
Gaussians of 1.0 kjoule/mol height and 5.0 width were deposited every 1000 steps.
A $\gamma$ value of 80 was used for all replicas.
Right panels: convergence of PT-WTE without exchanges (top) and of standard PT (bottom).}
\label{HIV_convergence}
\end{figure}

\begin{figure}[!h]
\begin{center}
\includegraphics[clip,width=1.0\linewidth]{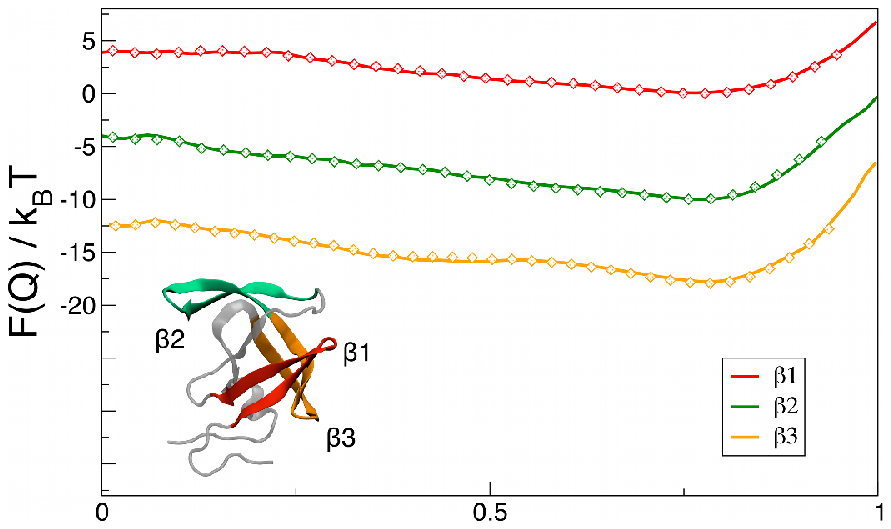}
\end{center}
\caption{G\={o}-model FES of three $\beta$-strand subunits of HIV--1 protease as a function of the native contacts at
$T=0.95$. The FES are obtained by reweighting the PT-WTE run (solid lines) and from an umbrella sampling calculation (points).}
\label{HIV_rew}
\end{figure}

As a further example of the power  of  PT-WTE, we show 
an application to the folding process of the monomer of HIV--1 protease. For this we use a 
G\={o}-model \cite{Clementi:2000p17845} which has 
a transition at $T_f\simeq80K$. For this reason, simulations
using straightforward PT give poor results unless the distribution of temperatures across $T_f$ is optimized \cite{Katzgraber:2006p17781}. 
In this example, we do not use the potential energy as CV, but
the variable on which the energy uniquely depends, namely the total number of native contacts between C$_\alpha$ atoms.
It is easy to show that in this case an expression equivalent to  Eq. \ref{WTE_delta} holds.
Simulations have been carried out using GROMACS \cite{Hess:2008p11450} and PLUMED \cite{Bonomi:2009p17107}. 
In this case $t_1/t_\gamma \simeq 66$. We also measure the speed-up in terms of MD steps 
needed to converge the free-energy difference between folded and unfolded state. 
In Fig.~\ref{HIV_convergence} we see that PT-WTE  converges
in less than $2.5 \cdot 10^7$ steps, while standard PT is still not converged after $2.4 \cdot 10^8$ steps.
We also show that allowing for replicas to exchange is crucial since 
WTE alone fails to converge in the simulation time.
As a further check we reconstruct the thermodynamics of three
relevant sub-units of HIV--1 protease (Fig.~\ref{HIV_rew}).
Comparing our results with  an umbrella sampling calculation that uses \emph{a posteriori}
 the PT-WTE bias, we find an excellent agreement.

In conclusion, we have shown that WTE can be profitably used as a biased ensemble to 
greatly enhance sampling speed especially when associated to parallel tempering. 
Properly designed WTE combines two properties that are useful in this respect. 
The fact that average values are not changed ensures  a significant overlap 
between the biased and unbiased ensemble facilitating the reconstruction of the latter. 
Yet the enhanced fluctuations favor exploring low probability regions and overcoming 
large barriers. 
Measuring the efficiency of this new method is a subtle question. We can claim
on the basis of Ref.~\cite{PhysRevLett.92.170601} that when it comes to reconstructing $N(U)$ we can
obtain an efficiency at least comparable to Wang-Landau. 
Furthermore, we have the additional bonus that we do not need extra calculations
or expensive reconstruction of multidimensional histograms to evaluate quantities
different from the energy or its fluctuations.
In this respect the fair comparison is with PT where we gain relative to Ref.~\cite{Katzgraber:2006p17781} as much as a 
factor of $\simeq100$ on the Ising model with $L=20$.
Much remains to be done to understand WTE properties and to optimize 
its performances. However, the very encouraging results obtained at these early 
stages suggest that a powerful method has been added to the literature and 
that exciting applications can be expected. 
Extension of the method in which additional CV are added to $U$ is straightforward and
will be explored in the near future.

We would like to thank Michele Ceriotti and Alessandro Barducci for fruitful
discussions. Calculations have been carried out on the BRUTUS cluster at ETH Zurich.

\end{document}